\documentclass{article}
\pdfoutput=1
\usepackage{graphicx}
\usepackage{listings}
\usepackage[english]{babel}
\usepackage{url}
\usepackage{tabularx}
\usepackage{multirow}
\usepackage{acronym}
\usepackage[colorlinks=false,pdftex]{hyperref}

\begin{document}
    

\title{Trends in Computer Network Modeling Towards the Future Internet}
\author{Jeroen van der Ham \and Mattijs Ghijsen \and Paola Grosso \and Cees de Laat\\
E-mail: \{vdham, m.ghijsen, p.grosso, delaat\}@uva.nl}%
\maketitle


\begin{abstract}
    This article provides a taxonomy of current and past network modeling efforts. In all these efforts over the last few years we see a trend towards not only describing the network, but connected devices as well. This is especially current given the many Future Internet projects, which are combining different models, and resources in order to provide complete virtual infrastructures to users.
    
        An important mechanism for managing complexity is the creation of an abstract model, a step which has been undertaken in computer networks too.  The fact that more and more devices are network capable, coupled with increasing popularity of the Internet, has made computer networks an important focus area for modeling. The large number of connected devices creates an increasing complexity which must be harnessed to keep the networks functioning.
    
    Over the years many different models for computer networks have been proposed, and used for different purposes. While for some time the community has moved away from the need of full topology exchange, this requirement resurfaced for optical networks. Subsequently, research on topology descriptions has seen a rise in the last few years. Many different models have been created and published, yet there is no publication that shows an overview of the different approaches.
\end{abstract}







\section{Introduction}%
\label{sec:introduction}

Communication networks, such as the Internet, play a fundamental role in modern societies and economies. It is nearly superfluous to remind anybody of the many changes that have occurred in the last twenty years since the invention of the World Wide Web and the wide adoption of the TCP/IP protocol suite.

Less known is that the role of networks is becoming even more central in emerging ICT architectures.  In these new infrastructures, which are labeled as \textit{Future Internet}, there is  a much more integrated operation of networking, computing and storage devices. All these components are being managed and monitored in a coordinated manner in order to deliver services to applications and end users.

One basic rule holds for both the current Internet and the upcoming Future Internet platforms: the design, planning, management and monitoring of the network rely on the knowledge of its topology. A network topology provides in fact information on the location of devices and on the connections between them; this information in turn gives a view of the physical and logical structure of the network. Topologies are expressed as \textit{network models}, and we use these two terms interchangeably in this article.

Topology information needs to be available to all devices within the network to operate properly, to external tools that act on the network and to applications that use the network. We see three main challenges for network models.
\begin{itemize}
	\item
\emph{Handling different abstraction levels:} From a devices perspective there is a wide range of topology details needed: at the edges of the network knowledge can be as minimal as knowing where the next hop is, while within the core devices require much more information. 
\item \emph{Managing multi-domain communication and path setup:}
External tools that operate on the network need to be aware of the network or to provide metadata of the network; monitoring tools require a comprehensive model to describe all relevant details of computer networks and the connections through them, while bandwidth-on-demand tools used in  circuit switched networks will only need to exchange some detail of network topology to be able to efficiently plan connections. 
\item \emph{Integration with computing-network-storage-planning services:} Once applications become more dependent on performance of the computer network they need more detailed models to be able to express their requirements, and closely monitor network performance.

\end{itemize}


In this article we provide an overview of some of the most used and well known network models. It is our intention to guide the reader through a historical journey that ultimately clarifies the need for new modeling approaches to support the Future Internet. To this end we first look at network descriptions in the history of the Internet in section~\ref{sec:history}. We then provide a categorisation of network models (section~\ref{sec:topo:categories}).

Following our model categorisation we present management models (section~\ref{sec:management_models}), monitoring models (section~\ref{sec:mon-models}) and generic models (section~\ref{sec:gen-models}). We also introduce the existing Future Internet model (section~\ref{sec:fi-models}). Section~\ref{sec:Discussion} provides an overview and discussion on the current state of network models research. We conclude the article with a summary and the upcoming research challenges in section~\ref{sec:conclusion}.




\section{Historical Role of Network Descriptions} 
\label{sec:history}
\begin{figure*}[hptb]
  \centering
    \includegraphics[width=0.9\textwidth]{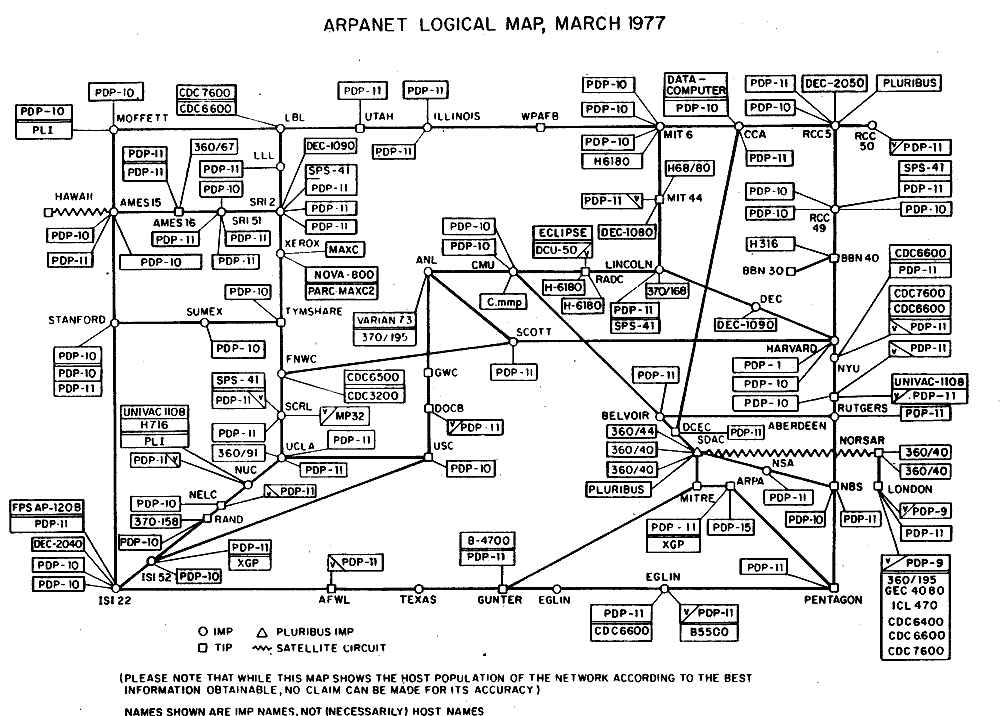}
  \caption{ARPANET logical map, March 1977, an example of early network models}
  \label{fig:Arpanet_logical_map,_march_1977}
\end{figure*}
Before delving into current network modelling efforts that aim to support the Future Internet, it is helpful to understand the role and evolution that network descriptions have had in the past years. 

We will show that networks have evolved from the original packet-switched architectures, to use optical circuit-switched designs to finally converge towards the Future Internet hybrid models, i.e. networks offering both packet and circuit switching services. We will also show that during this evolution there is one constant requirement that has not changed: the need to exchange information about the network topology. For packet-switched networks topology information is needed for the operation of routing protocols, for circuit-switched and hybrid networks it is required for the creation of dedicated connections among end-points.

\subsection*{Packet-switched networks}
Topology descriptions have been used to support computer networking activities since the start of the Internet. The most commonly used technique to capture a network topology is of course a graphical representation. One of the obvious drawback of this method is that it does not scale well as the network becomes larger, making automated tools necessary. Fig.~\ref{fig:Arpanet_logical_map,_march_1977} shows an early representation of the ARPANET\cite{arpanet}. This network started out with just four nodes in 1969, but quickly grew larger. The figure shows a large network with many devices and connections which is hard for humans to grasp in its entirety.

The ARPANET originally used Interface Message Processors (IMPs) to route messages through the network\cite{Heart1970}. These IMPs performed regular delay measurements to all destinations, and then broadcasted the result. These results were combined and then stored to function as a sort of distance-vector protocol. Over the years the routing between IMPs was gradually improved, until in 1983 the ARPANET switched over to TCP/IP.

Research on TCP/IP had already been going on during the seventies on several test networks\cite{Cerf1974}. During this time the \acl{RIP}\cite{rfc1058} was also developed, implementing a distance-vector protocol. Distance-vector protocols form an abstract view of the network, using the distance and general direction as a way to select the forwarding interface. Similar to this is the path-vector \acl{BGP} \cite{rfc1105,rfc4274} which rely on operator defined paths in the network, serving most backbone networks in the Internet.
 These protocols no longer need a complete picture of the network. Instead, each router has a (different) aggregated view of the network, gathered from exchanging aggregated information with others.

During the late 70s and early 80s several different link-state routing protocols were developed, among them \acs{IS-IS} and \acs{OSPF}\cite{McQuillan1979,Perlman_1991}. Link-state routing protocols broadcast messages containing the states of links, and where they directly connect to. Traditionally this broadcasting is limited to smaller areas, and not the whole network. Within an area, all routers do form a complete view of the topology, and use this to calculate the shortest path tree. Link-state networks are mostly used in local networks.

When relying on distance vector protocols full topology distribution is no longer a requirement. However, having a full network description available is still needed by link-state protocols and it can in general still be helpful in monitoring or problem detection. In both cases the transfer of information regarding the topology is done directly by the network nodes.

\subsection*{Circuit-switched and hybrid networks}
The methods to derive knowledge on the network topology we just described have been driven by the routing protocols, and as such they are only applicable to packet-switched networks. They are not very useful in the context of circuit-switched or hybrid networks. Models better suited for these latter situations have emerged in the past years.

For circuit-switched networks  full topology distribution is still required. In order to send data from a source to a destination in a circuit-switched network, a circuit must be configured. In telephony networks dynamic provisioning is achieved by using strict addressing, aggregated static routes, and large capacity\cite{ss7-routing}. For circuits in data networks this is not feasible, since there is no strict numbering plan, and the overall capacity compared to the circuits is not that great.

Asynchronous Transfer Mode tried to merge the world of circuit-switched networking with packet-switched networking. There the Private Network-to-Network Interface \cite{pnnispec} was used to relay topology information, and also included some ideas on topology aggregations. In the end, ATM never became very widely adopted, and is not currently in mainstream use.

\acs{GMPLS} with its Path Computation Element\cite{dasgupta07pce} takes a different approach for inter-domain path computation. Instead of sharing topology information, every request in the network is broadcast to peers. The route of the request is recorded and replied along the same path to implement a circuit reservation request. While technically feasible, this approach poses problems as the number of requests goes up. While GMPLS is implemted intra-domain, we have not seen inter-domain deployments.

A different approach is seen in hybrid networking\cite{rationale-optical}. Many research and educational networks are currently offering circuits on their own network, and recently also started experimenting with inter-domain circuits\cite{nsiv1}. Here the topology of a domain needs to be exported in full or in an abstracted way to the neighboring domains. The representation of the network needs to be consistent and agreed upon, such that inter-domain circuit provisioning tools can take decisions on how to engineer a circuit.

While the ARPANET and Internet have moved away from the need of full topology exchange, the need for topology description and exchange has risen again for optical networks. Subsequently, research on topology descriptions has also seen a rise in the last few years. Many different models have been created and published, yet there is no publication that shows an overview of the different approaches.


\section{Topology Categories} 
\label{sec:topo:categories}
\begin{figure}[htbp]
  \centering
    \includegraphics[width=.5\columnwidth]{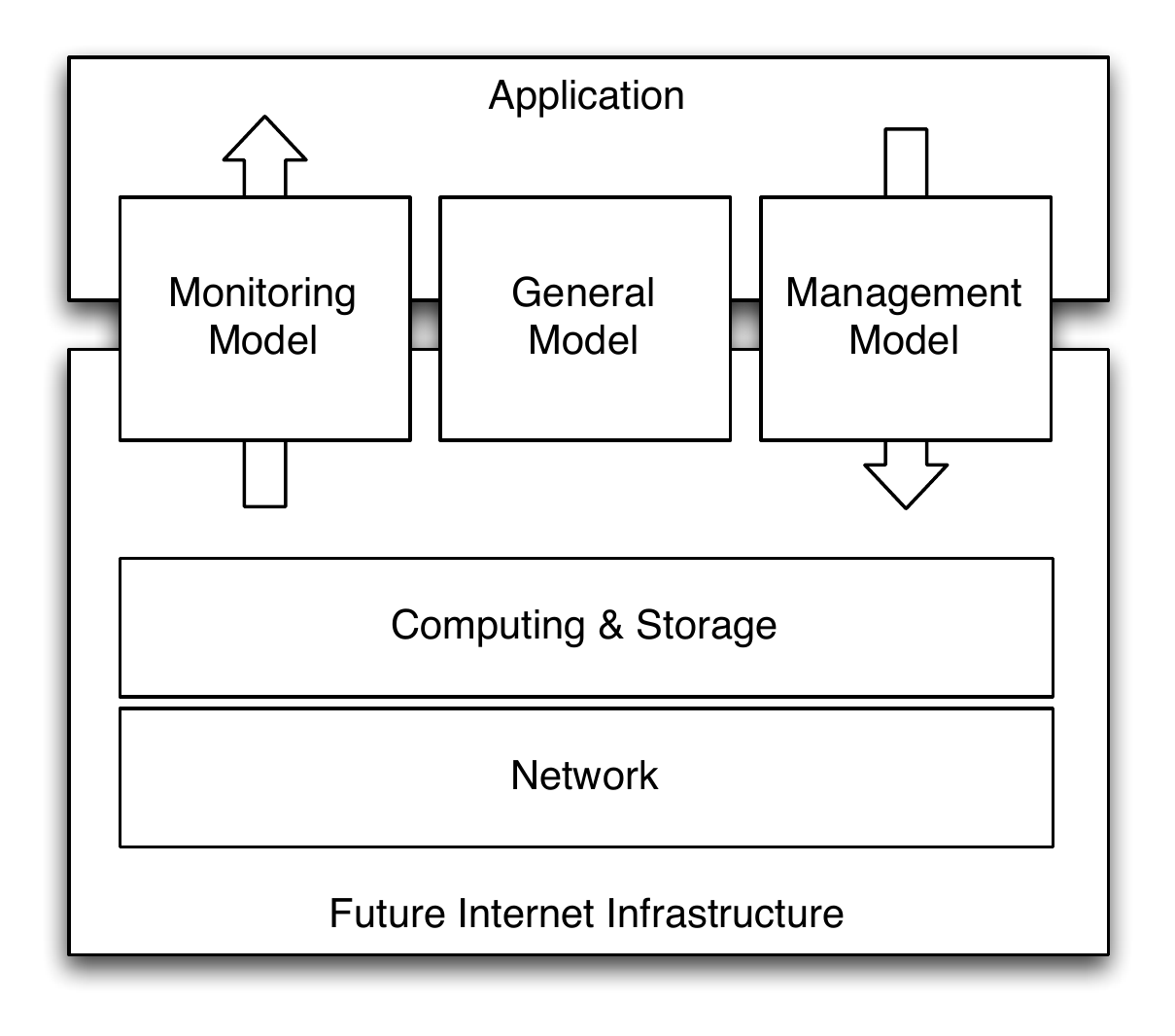}
  	\caption{Computing Infrastructure Models}
  	\label{fig:computing-infrastructure}
\end{figure}

The historical perspective we just gave provides a sense of why models are needed, and how they have been used concretely. But it is also useful to categorise the various models in a more general way. We can, in fact, analyse and compare different computing models suitable for Future Internet infrastructures based on the following three features:
\begin{enumerate} 
	\item their purpose from an application perspective,
	\item the range of infrastructure layers covered by the models,
	\item the functional scope covered by the models.
\end{enumerate} 
An overview of these features and how they relate to each other is shown in Figure~\ref{fig:computing-infrastructure}, where we provide two main blocks, i.e application and future internet infrastructures, and we position models in them according to their characteristics. 

From an application perspective, we distinguish between three different models in terms of the type of application they support. 
\begin{itemize}

    \item \emph{Management} models are used in network-management applications or to restrict actions that can be taken on a network. 

    \item \emph{Monitoring} models are used for external applications to describe the dynamic aspects of a computing infrastructure. 

    \item \emph{General} infrastructure models are used for applications that require a static view of computing infrastructures. 

\end{itemize}

When starting from the infrastructure perspective, different models cover a different range of layers in the infrastructure. In this paper we distinguish between models that focus on a single technology layer of the infrastructure and models that cover multiple layers of technology. We also identify two different functional aspects of a Future Internet infrastructure that can be covered by a model. Most of the models discussed are focused on the network infrastructure that connects the different resources in the computing infrastructure while other models also include computing and storage capabilities of the Future Internet infrastructure.

Besides the content, we will also take the modeling approach into account for comparing and analyzing different models. For this purpose we identify the following types:
\begin{enumerate}
	\item byte format, used in communication protocols and aimed at compact descriptions;
	\item database schema, used to describe the content of the database in which the instances of the infrastructure are stored;
	\item Unified Modeling Language, used to describe the classes and relations in an object oriented model;
	\item Extensible Markup Language (XML), used to provide a schema for the model and syntax that is application and programming language independent; 
	\item Semantic-Web based models, i.e. models based on the \acl{RDF} or the \acl{OWL}, used to provide semantic models of future internet infrastructures.
\end{enumerate}

\begin{table*}[htpb]
	\centering
	\caption{Overview of model characteristics.}
	\label{table:modeloverview}
	\begin{tabular}{|l|l|l|l|l|l|}
\hline
                                               \textbf{Model} & \textbf{Main Purpose} &    \textbf{Scope} &   \textbf{Type} & \textbf{Standard Organization} & \textbf{References}\\
\hline
\hline
                           \acs{SNMP} &            Management &           Network &       DB schema &                     \acs{IETF} & \cite{rfc3410}\cite{snmp-book}\\
\hline
                                NetConf &            Management &           Network &             XML &                     \acs{IETF} & \cite{rfc4741}\cite{rfc6241}\\
\hline
      \acs{OSPF}(-TE)/\acs{GMPLS} &            Management &           Network &     byte format &                     \acs{IETF} & \cite{rfc3945} \cite{farrel-gmpls} \\
\hline
                                          \acs{CIM} &            Management &  Network + Comp \& Storage &                      UML + XML & \acs{DMTF} & \cite{cim}\\
\hline
                                    \acs{DEN-ng} &            Management &  Network + Comp \& Storage &                            UML & \acs{DMTF} & \cite{den-ng}\\
\hline
              perfSONAR/\acs{NMC} &            Monitoring &           Network &             XML &                      \acs{OGF} & \cite{perfsonar}\cite{perfsonar2}\\
\hline
                                 cNIS &            Monitoring &           Network &       DB schema &                              - & \cite{cnis-db}\cite{autobahn1}\\
\hline
                                    \acs{MOMENT} &            Monitoring &         Network + Comp. \& Storage &                      \acs{OWL} &          - & \cite{moment}\\
\hline
            G.805/G.809/G.800 &               General &           Network &            None &                            \acs{ITU}& \cite{g805,g809,g800,g805-intro}\\
\hline
\acs{NDL} &               General &           Network &       \acs{RDF} &                              -  & \cite{vdham06ndl}\cite{g805-network-model}\cite{vanderHam200885}\\
\hline
                                   \acs{NML} &               General &           Network & XML + \acs{OWL} &                      \acs{OGF} & \cite{nml-schema} \\
\hline
                                  \acs{RSpec} &               Request &            Comp \&         Network &                            XML &          - & \cite{url:rspec}\\
\hline
                                        \acs{VxDL} &               Request &            Comp \&         Network &                            XML &          -& \cite{vxdl}\\
\hline
                              \acs{NDL-OWL} &               General &  Network + Comp \&         Storage &                            OWL &          - & \cite{orca1,orca2}\\
\hline
     NOVI/GEYSERS/INDL &               General &  Network + Comp \&         Storage &                      \acs{OWL} &          - & \cite{VanderHam2011}\cite{NOVID22}\cite{indl}\cite{indl2}\\
\hline
	\end{tabular}
\end{table*}

In Table~\ref{table:modeloverview} we provide an overview of the models and their main characteristics discussed in the following sections.

\section{Management Models}\label{sec:management_models}

Network management has used several different information models over the years, and newer models are being proposed. These models are mainly used for management of devices, or in protocols to exchange necessary topology information. They are generally aimed at specific applications, the information expressed in the protocols is not meant to be generically available nor extensible.

\subsection{SNMP} 
\label{par:snmp}

The Simple Network Management Protocol\footnote{Technically, the information model is formed by the MIBs, Management Information Bases, and SNMP denotes the whole set: protocol, information and data model.}\cite{rfc3410,snmp-book} is a set of standards describing a protocol, a database schema, and data objects. The whole suite was originally created as a way of both monitoring and managing network resources. In current networks it is mainly used for monitoring purposes.

Diagnostic, performance and configuration information of network devices can be retrieved from the \ac{MIB} of devices using \ac{SNMP} messages. The \ac{MIB} is a tree of name -- value pairs, which can be requested and changed. The values are restricted to three different types of datatypes: integer, string and sequence of datatypes. A large part of the MIB tree is standardised, but vendors also have their own private part of the tree. This vendor space is used to store most configuration and performance data of their devices in a proprietary format. Virtually all networking devices support \ac{SNMP}, with different levels of detail in their \ac{MIB}.

The network description provided by \ac{SNMP} is distributed over the devices. Depending on the layer the device is operating on, it may have a pointer (address or identifier) to its neighbours on that layer. A view of the whole topology can be created by combining the information gathered from all the devices.

\subsection{NetConf} 
\label{par:netconf}

The Internet Engineering Task Force has recently worked to replace \ac{SNMP} with a new standard, NetConf\cite{rfc4741}\cite{rfc6241}. While \ac{SNMP} uses its own protocol and only allows for three data-types, NetConf uses XML, allowing for many more data-types. NetConf defines a way of transporting monitoring data and change requests over a small set of existing protocols. NetConf is aimed at distributing diagnostic, performance and configuration information, but also for managing devices. NetConf is currently being introduced in networking devices.

As NetConf follows similar principles as its predecessor, the network description provided by NetConf is similarly distributed over the managed devices. Each device will have information about the neighbour it connects to on the layer it operates on. The network topology can be created by combining the information of the devices in the network.

\subsection{GMPLS} 
\label{par:gmpls}

\acs{GMPLS}, Generalized Multi-Protocol Label Switching\cite{rfc3945}\cite{farrel-gmpls}, is a protocol suite developed by the \acs{IETF} for the provisioning and management of label-switched paths through multi-technology networks. It provides a unified control and management plane for the management of multi-layer networks. Networking devices use the \ac{OSPF-TE} protocol to exchange topology data with their neighbours. Devices broadcast the received topology data to their other neighbours, so that in the end all the devices in the domain have the same view of the network topology. 

The topology data in \ac{OSPF-TE} is exchanged in Link State Announcements packets inside network domains. The topology data contained therein is encoded in a compact byte format, using specifically defined header fields and Type-Length-Value containers. This format is designed to be easy to process and store for participating network devices, but it is hard to export to external applications. The message format is somewhat extensible, there is specific room for other applications to add data to the messages. The data must fit in the Type-Length-Value container, and can be processed by agents participating or listening to the \ac{OSPF-TE} process.

Since \ac{OSPF-TE} is only used intra-domain, there is no inter-domain exchange of messages or information. In order to allow for inter-domain provisioning, the Path Computation Element architecture \cite{rfc4655} has been defined. \ac{GMPLS} operators have expressed a desire to keep network topology data confidential, so the path computation architecture works by broadcasting requests, rather than by distributing topology information\cite{dasgupta07pce}\cite{rfc5520}.


\subsection{CIM and DEN-ng} 
\label{sub:cim}

The \ac{CIM}\cite{cim} is a network device information model commonly used in enterprise settings. \ac{CIM} is developed by the \acl{DMTF}\cite{dmtf} and it is an object-oriented information model described using the Unified Modeling Language. This information model captures descriptions of computer systems, operating systems, networks and other related diagnostic information. \ac{CIM} is a very broad and complex model, the current UML schemata of the network model span over 40 pages, the total model is over 200 pages.

A mapping from \ac{CIM} to XML is also defined, which is mainly used in Web-Based Enterprise Management. This is mainly implemented in enterprise-oriented computing equipment, and operating systems such as Windows and Solaris.
The \ac{CIM} model is highly expressive, and is still actively developped. There have been many significant changes in the infrastructure part of \ac{CIM} over the past two years, both introducing new elements, as well as deprecating or changing existing elements. The \ac{CIM} model is capable of capturing the complete physical setup, and almost everything with regards to the configuration of devices. The model is capable of capturing the information with a very high level of detail, yet provides almost no abstraction layer above this, making it very hard to reason generically using this model.

A successor to \ac{CIM} is the \ac{DEN-ng} model, Directory Enabled Networking -- next generation\cite{den-ng}, which extends the \ac{CIM} model also with description of business rules. The idea behind the model is that with the right software, the business rules combined with the capabilities of the devices can be automatically transformed into configurations of firewalls, user restrictions, et cetera. This requires that all configuration management is managed centrally, or at least by the same tools.

\section{Monitoring Models} 
\label{sec:mon-models}

The previous section provided an overview of management models, which are usually aimed at specific tools for network and device management. Many communities like to provide more generic access to monitoring data, so monitoring models have been created. These models can take output from different tools and combine them into a single model.

\subsection{perfSONAR / NM and NMC} 
\label{sub:perfsonar_nm}

An early model for network topology description is the perfSONAR\cite{perfsonar}\cite{perfsonar2} model. perfSONAR is a network monitoring architecture. It stores data from different measurement tools which are then made available publicly. This is particularly intended for inter-domain network connection debugging\cite{perfsonar-inter-domain}.

The perfSONAR architecture has been implemented by different partners, providing two different, compatible implementations. The model has later been brought to the Open Grid Forum (then Global Grid Forum)\cite{ogf} for standardisation. This resulted in the \ac{NM-WG}\cite{nmwg} which produced a standardised schema in 2009\cite{nmwg-schema}.

The \ac{NM-WG} schema contains a base schema to describe network measurement tools, and their results. There is also a time schema to accurately describe time values in these measurements. Of particular interest here is the topology schema, which provides a basic representation of network topologies using hierarchical constructs in XML. This schema allows for a simple description of domains, nodes, ports and their connections.

This schema is also used in the Inter-Domain Controller Protocol\cite{idcp}, which is currently in use in many circuit provisioning tools, e.g. OSCARS\cite{guok06oscars}. The OSCARS tool allows users to make circuit requests for the Energy Sciences Network (ESnet\cite{esnet}), and has also been implemented on the Internet2 ION network\cite{i2-ion}.

The \acl{NMC} has currently taken over the activities of the activities of the \ac{NM-WG} and is continuing development of the measurements schema. The topology schema development has moved to the \acl{NML}, which we discuss later.

\subsection{cNIS and AutoBAHN} 
\label{sub:cnis}

cNIS is the network topology description format for G\'EANT network\cite{url:geant} and is used as basis for the AutoBAHN\cite{autobahn1} bandwidth on demand system. The data model is implemented in a database schema\cite{cnis-db}. This schema includes fixed descriptions of a set of layers used in the G\'EANT network, such as Ethernet, and MPLS.

The AutoBAHN bandwidth on demand system at first started with the cNIS, but later extended it towards their own model\cite{autobahn2}. The AutoBAHN system uses a Domain Manager which maintains the local topology. This Domain Manager does automatic topology aggregation before exporting a topology to the Inter-Domain Manager. Interestingly, the Inter-Domain Manager uses extensible OSPF messages to exchange inter-domain topology information.

The Stitching Framework\cite{sfw} is also a G\'EANT activity, and it describes a framework for `stitching' together different technologies in bandwidth-on-demand systems in a multi-domain and multi-layer environment. It provides a framework to define the required information for creating connections across multi-domain multi-layer networks. The Stitching Framework has been integrated into the latest version of cNIS where it can stitch together the technologies defined there. It should be noted that the Stitching Framework is built generically, and could also be applied to other more expressive models.

\subsection{Monitoring and Measurement Ontology} 
\label{sub:moment}
The perfSONAR and \ac{NM-WG} work served as an important inspiration for the \ac{MOMENT} developed by ELTE\cite{moment}. This ontology has taken the initial concepts from \ac{NM-WG} and implemented them into an \ac{OWL}-based ontology. This ontology is mostly aimed at measurement tools and results, which using their \ac{OWL} ontology, can both be expressed in great detail.

The ontology allows an application to describe the exact circumstances of a measurement. For example that a \texttt{traceroute} command was performed at a certain time, the parameters of that command, a description of the network at that time, and the results of the command itself. These kinds of measurements can then be recorded in a database, where they can be easily correlated and analyzed using the generic description of the data.

The \ac{MOMENT} ontology has served as a way of describing data for the ETOMIC\cite{Matray07etomic} infrastructure. This infrastructure consists of several nodes together forming a network measurement virtual observatory. The OWL-based ontology then makes it possible to easily share and reuse measurement data with others.

The experiences of the \ac{MOMENT} ontology have been used also in the development of the NOVI monitoring ontology.


\section{General Models}
\label{sec:gen-models}

In the previous sections described management and monitoring models, which are aimed at management and monitoring applications respectively. Another category is the set of general models, which aim to provide a more general description of the network topology so that other applications can use them.

\subsection{G.805, G.809 and G.800} 
\label{sub:g_805_and_g_809}

A very generic set of models are the network models defined by the \ac{ITU}. These models are theoretical models, in the sense that they have no explicit data model defined for them. However they are important to discuss here as they have identified and defined important terminology for network topology description, especially concerning multi-layer networks.

In 2000 the \ac{ITU} published the G.805 network model\cite{g805}. This model allows the description of all kinds of transport networks, and especially different layers and adaptations in that network. It is a very comprehensive, but also complex model. A more readable introduction is available\cite{g805-intro}.
The G.805 model allows the modelling of circuit-switched networks, and in 2003 the model was extended in G.809\cite{g809} to also model connection-less networks. Then in 2007 these models were combined, along with some others into G.800: `Unified functional architecture of transport networks'\cite{g800}.

These models are very extensive and generic, allowing to describe any kind of existing network, but also future network technologies. The models have identified some fundamental concepts, such as:

\begin{itemize}
    \item \emph{Layers} is defined as the set of connection points of the same technology,
    \item \emph{Adaptations} are the functions performed on data to transform it from one layer to another,
    \item \emph{Labels} identify different flows of data in a Layer.
\end{itemize}

So as a simple example, VLAN tagged traffic is a specific Layer, the adding of a VLAN tag to a packet is an Adaptation, and the VLAN tag is used to identify a data flow among the other traffic.

However, G.805, G.809 and G.800 are only graphical models, there is no data model underlying these information models, making them hard to use in practice. The models do provide a very fundamental theoretical groundwork, which is why \acs{NDL} and \acs{NML} have taken it as a source of inspiration.

\subsection{Network Description Language} 
\label{sub:network_description_language}

In 2006 the University of Amsterdam published a method of using RDF to describe networks\cite{vdham06ndl}, called the \ac{NDL}. This uses a simple model to describe devices, interfaces and their connections. The descriptions would then be available to applications in a standard format. The initial idea was also to apply the distributed description capability of the semantic web, similar to the Friend of a Friend network\cite{url:foaf}. This allows networks to independently describe their network topologies and link them together so that they together form a global description of the network.

The initial model of \ac{NDL} (v1) was simple, and in some ways similar to the model used by PerfSonar, but implemented in \ac{RDF}. Using ideas from G.805 we extended \ac{NDL} to version 2, which describes multi-layer networks generically\cite{g805-network-model}\cite{vanderHam200885}. This model introduces a notation for the G.805 concepts of Layers, Adaptations and Labels. This allows for descriptions of any kind of network topology, ranging from physical networks to completely virtualised networks, and also the relations between those network layers.

\ac{NDL} has been used as one of the models on which the Network Markup Language is based, and also heavily influenced the design of the \acs{NOVI} and \acs{GEYSERS} information models.


\subsection{Network Markup Language} 
\label{sub:network_markup_language}

During 2007 efforts have been combined from PerfSonar, \ac{NM-WG} and \ac{NDL} to create a standard network topology information model. A new working-group was formed at the \acs{OGF} called the \acl{NML}\cite{url:nml}. This group aims to create a generic network model that can be used for describing measurements, monitoring, describing topologies, and also requests.

The \ac{NML} schema describes networks using uni-directional constructs. The unidirectional Port objects can be connected together, externally through Links or internally through a Node's Crossconnect. The model also includes the capability of describing multi-layer networks based on the ideas from G.805 and \ac{NDL} as described earlier.
The unidirectional model causes the network model to be very verbose, however this allows the model to be more generic, as a unidirectional model can describe bidirectional networks, but vice versa this is not possible.

The standardisation process has recently resulted in the publication of the first \ac{NML} base schema\cite{nml-schema}. To support different applications, \ac{NML} has two different data models, one in XML and one in \ac{OWL}.


\section{Future Internet Models}
\label{sec:fi-models}
In recent years several initiatives have started to work on so-called Future Internet platforms. Examples are the \acs{GENI}\cite{url:geni} initiative in the United States, and the \acs{FIRE}\cite{url:fire} initiative in the Europe. From these several different projects have started, which we discuss below.

\subsubsection{RSpec \& RSpec v2} 
\label{sub:rspec_&_rspec_v2}

The \acs{GENI} project\cite{url:geni} in the United States has been working on very large distributed virtualization infrastructures, such as PlanetLab\cite{planetlab,url:planetlab}, and ProtoGENI\cite{protogeni}. These testbeds contain nodes distributed over different locations, connected to the Internet, where users can request virtual machines and conduct network experiments.

Initially PlanetLab developed the \ac{SFA} format to provide infrastructure and request descriptions. The first version of this format have been defined in \ac{RSpec}\cite{url:rspec}. This later evolved into ProtoGENI \ac{RSpec} v2\cite{url:SFAv2}, which has been chosen as the standard interchange format for PlanetLab, and all other \ac{GENI} platforms. 

The RSpec v2 format is a simple XML based format geared towards the specific use in virtual environments. It allows platforms and users to describe nodes, their virtualisation properties, and a very limited form of network connectivity. The format works very well with PlanetLab and compatible systems, but it is very hard to use when describing any other kind of network or infrastructure.

\subsubsection{Virtual private eXecution infrastructure Description Language} 
\label{sub:virtual_private_execution_infrastructure_description_language}

The \ac{VxDL} has been developed by INRIA and Lyatiss\cite{vxdl,lyatiss}. \ac{VxDL} uses an XML syntax to express infrastructure requests in varying levels of detail. Such a request consists of four parts: a general description, a description of non-network resources, a network topology, and the time interval for this reservation.

\ac{VxDL} is used in GRID5000\cite{grid5000}, the \acs{GEYSERS} project (see section~\ref{sub:geysers_and_novi}, as well as a commercial product developed by Lyatiss.

\subsection{Network Description Language OWL} 
\label{sub:network_description_language_owl}
RENCI\cite{url:renci}, a \ac{GENI} participant, has also built an infrastructure, called ORCA-BEN\cite{orca1,orca2}. This infrastructure contains several locations with virtualisation capabilities, and a completely controllable optical network. In order to control and manage this they have extended NDLv2 to the OWL syntax, creating \ac{NDL-OWL}. This also extends \ac{NDL} with more virtualisation and service description features to describe their infrastructure. These descriptions are then used in the client software to describe requests, but also in the management software to match the requests with the available infrastructure.

The development of \ac{NDL-OWL} and ORCA-BEN has been performed in the context of the \ac{GENI} project, which means that ORCA-BEN is able to communicate with other \ac{GENI} platforms, including platforms speaking \ac{RSpec} v2. \ac{NDL-OWL} is thus a superset of \ac{RSpec} v2.

\subsection{NOVI, GEYSERS and INDL} 
\label{sub:geysers_and_novi}

The \acs{NOVI} project aims to federate Future Internet platforms and one of the challenges of the \acs{NOVI} Information Model is to interact with different platforms\cite{VanderHam2011,NOVID22}. Using \ac{NML} in the information model provides the basis for interaction between \acs{NOVI} and the FEDERICA and PlanetLab platforms. Not only does the information model have to map to concepts used in these platforms, it also needs to be able to accommodate interaction with other platforms that may be added to the federation in the future.
By adding concepts from the \ac{MOMENT} ontology also to the \acs{NOVI} ontologies, users can easily use monitoring tools and data to get a comprehensive view of their requested infrastructure. The \acs{NOVI} ontology suite allows a complete semantic description of a Future Internet federation. \acs{NOVI} has ontologies for the infrastructure, but also for monitoring tools and results, as well as policy aspects and rules. Of special interest in the \acs{NOVI} model is the \emph{unit} ontology, which generically describes the units used for capacity, measurements, et cetera.

One of the key innovations of \acs{GEYSERS} is to enable virtualisation of optical infrastructures. The \acs{GEYSERS} Information Modeling Framework (IMF), is currently under development to provide an information model for the Logical Infrastructure Composition Layer~\cite{Garcia-Espin2012}. This layer is the element responsible of managing physical resource virtualisation and composing Virtual Infrastructures. These are then offered as a service within the \acs{GEYSERS} architecture. 

The information models in both \acs{NOVI} and \acs{GEYSERS} are used to both describe the infrastructure and also to allow users to express requests. Once an infrastructure request is handled by either system, the result is also described in the same information model and made available to the user. This description can then also be used to correlate data from the active monitoring tools.

These platforms show that infrastructure provisioning is a complex interplay of different hardware and software tools, which benefits greatly from having an interoperable semantic model to exchange information. These models combine many aspects of the previous models, providing users with a single semantically compatible model for describing requests, physical and virtual infrastructure, as well as directly related monitoring information.

The \ac{INDL}\cite{indl,indl2} is an evolution of the \acl{NDL}, combined with the experiences in \acs{NOVI} and \acs{GEYSERS}. We have taken the general model from \ac{NML}, and added capabilities to describe the virtualisation of nodes and infrastructure. The model is actually not that different from the model in \acs{NOVI} and \acs{GEYSERS}, but provides a more reusable model available for other Future Internet platforms.




\section{Discussion}
\label{sec:Discussion}


\begin{figure*}[!htbp]
   \centering
       \includegraphics[width=.8\textwidth]{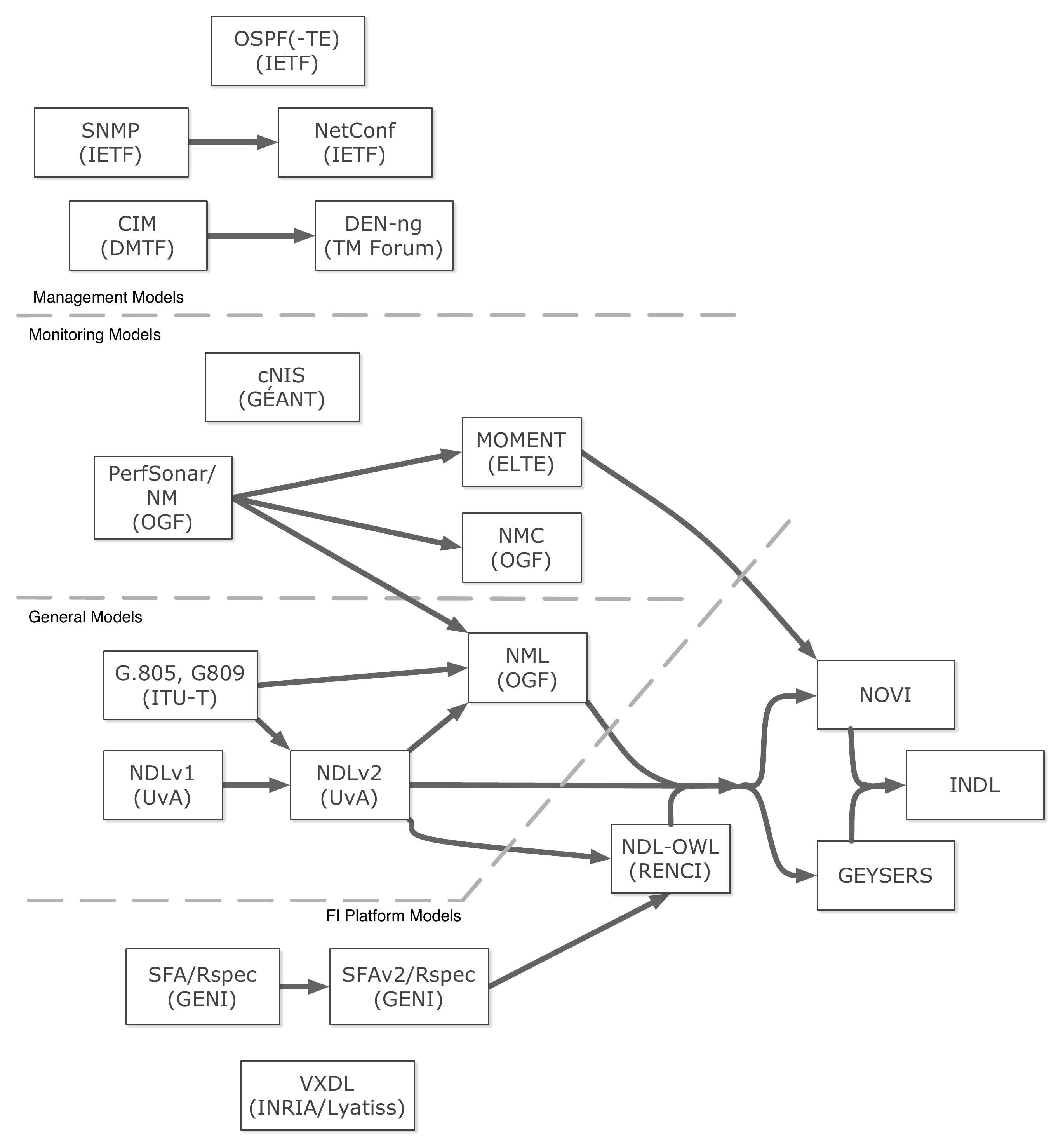}
   \caption{An overview of different information models}
   \label{fig:InformationModels}
\end{figure*}

This article presented an overview of the current state-of-the-art of network description models, with the goal to show how these models are suitable to the needs of Future Internet platforms. Figure~\ref{fig:InformationModels} shows an overview of the described information models and how they have influenced each other. This figure groups the models by intended usage: at the top of the figure we have the models more related to  management; we then show the monitoring models, and below them  the general models. The Future Internet models which combine the ideas of the monitoring and general models to form a complete ontology for future infrastructures are on the bottom and right side of the figure.

The information models described in section~\ref{sec:management_models} are aimed at describing purely functional topology and diagnostic information, making these management models. For example the \ac{GMPLS} information model is aimed switches and routers. The data model is designed for compactness and is therefore not easy for other applications to understand, nor is it human readable.
\ac{CIM} and \ac{DEN-ng} are also management models, albeit at a higher level, combining all the information of different low-level management models. This creates an aggregated management model at the enterprise level. The information models in these categories are both aimed at management, informing the direct operators of those machines.
These management models are aimed specifically at a single task, which they perform very well. The models are used in isolated contexts and domains, and the models are not generic enough to be used in applications not specifically aimed at these contexts. Most of the other information models described in this article have some form of an XML data model and are thus more generally usable. 

The monitoring models, PerfSonar/\ac{NM-WG}, cNIS and \ac{MOMENT}, have been defined specifically to capture data from many different tools, and store and share them in a generic way. These monitoring models are targeted at capturing monitoring information, network measurement data along with topology data of those measurements. Unlike the  management models, the measurement models aim to make the data as portable as possible so that different tools and applications can interpret the data, instead of a single management application. The network topology description elements of these models support the description of results, and are not that advanced in describing different technologies, or the dynamics between the technologies.

The general models are aimed specifically at describing network topologies. The initial model of \ac{NDL} was also not capable of describing multi-layer networks, but this changed due the influence of the \acs{ITU} G.805, G.809 and G.800 models. The \acs{ITU} models have very clearly identified and extensively defined a terminology for multi-layer networks. Using the generic (de)adaptation and labelling concepts it becomes possible to describe any kind of technology, without being dependent on a predefined notation for that technology. The \ac{NDL}, and \ac{NML} models aim at generic network descriptions that can be extended or embedded in other models. The intention of the generic models is to provide applications using the data enough information to act on the network, either by provisioning circuits or by adapting the applications behavior to the capabilities of the network.

The general models have been very influential in the creation of most models of the Future Internet. The initial models, \ac{SFA} and \ac{VxDL}, created for the Future Internet have been limited models to allow users to easily describe their requests for virtual infrastructures. The later Future Internet models, \ac{NDL-OWL}, \acs{NOVI}, \acs{GEYSERS} and \ac{INDL}, have built on both these simpler request-like models, as well as the general models to support the management of the Future Internet testbeds. This support is both for users in clearly defining their requests and the resulting topology. But the model also supports the management of the testbed to describe in a single model the physical resources, as well as the reserved virtual infrastructures.

The semantic web nature of the general models allow them to be easily incorporated in other models. Which is what we see happening somewhat in the NDL-OWL model, but even more so in the \acs{NOVI} and \acs{GEYSERS} models. The have taken the basic network models of \ac{NDL}/\ac{NDL-OWL}/\ac{NML} and extended these ideas towards virtual infrastructures and also adapted the request models to form a single information model for Future Internet infrastructures. The \acs{NOVI} model takes this another step further by also integrating the \ac{MOMENT} monitoring and measurement ontology, forming a complete semantic network model.

\section{Conclusion} 
\label{sec:conclusion}

Our article documents a clear evolution in the modeling of networks and infrastructure toward supporting Future Internet operations.

On one hand we have shown that management models have changed less, given they are all aimed at specific applications, and target very specific use-cases or tools. 
The hardware or chosen management software limits the choice for an information model in this case.

On the other hand, monitoring, general, and Future Internet models have all evolved significantly. The evolution we have documented shows that from several different initiatives at first, there has been a convergence on the newly defined \acl{NML} standard. Many of the models were of direct influence to \ac{NML}, so the standard is suitable for use in monitoring, provisioning as well as request modelling. The Future Internet models we are interested in have  taken \ac{NML} as their base model and extended it where necessary to describe resources beyond the network topologies. These extensions are also again converging in an extended model, \ac{INDL}.

\subsection{Challenges for Network Models}
Computer networks have become complex systems over the years and interactions with the network, especially circuit switched networks, should not be taken for granted. Our overview of the different models we presented demonstrates  that creating an information model for computer networks is not a simple feat. This is even more true for Future Internet platforms: there, networking is becoming more and more ubiquitous and more integrated in the  computing-storage fabric, making the management of computer networks a much more difficult task. 


We have identified three challenging areas for network models in the coming years:

\begin{itemize}
	\item handling abstraction levels appropriately;
	\item managing multi-domain communication and path setups;
	\item integration with computing-networking-storage planning services.
\end{itemize}

In 35 years we have moved from a situation where the entire Internet could be captured in a single figure (see Figure~\ref{fig:Arpanet_logical_map,_march_1977}) to a situation where we are running out of IPv4 address, with many more devices hidden behind NAT solutions. Network management has no choice but to move with this pace, requiring higher abstraction levels. Network information models are a necessary prerequisite for creating these abstraction levels. Current models do not adequately handle different abstraction levels in the same models.

Network descriptions are important in supporting path selection tools. Consider the architecture described by Lehman et al.\cite{lehman11multilayer} which points to the fact that an interoperable inter-domain topology description is necessary in order to allow path selection for multi-domain multi-layer circuit-based networking. Path selection in single layer networks is trivial, however in multi-layer networks it is much harder, and often NP-complete\cite{Kuipers200978}. The generic way of describing network technologies enabled by the abstract models of G.805 and G.809 makes it possible to create generic path selection algorithms which will be able to handle many if not all existing and future network technologies. The way that network topologies are represented are an important factor in supporting the path selection process. 

The problem of multi-layer path selection has many similarities with matching requests with (virtual) infrastructures. The nodes and services that are part of the request can be seen as special kinds of links connected to the network, similar to multi-layer network requests. By using generic models the application can choose to solve this problem directly, or it can choose to carve the problem up and delegate subproblems to the relevant planning services. This will lead the way towards a complete Future Internet infrastructure.


\section*{Acknowledgments}
This research was financially supported by SURFnet in the GigaPort-NG Research on Networks project and the Dutch national program COMMIT. 

\section*{List of Abbreviations}
\begin{acronym}[GEYSERS]
\acro{BGP}{Border Gateway Protocol}
\acro{CIM}{Common Information Model}
\acro{DEN-ng}{Directory Enabled Networking - next generation}
\acro{DMTF}{Distributed Management Task Force}
\acro{FIRE}{Future Internet Research and Experimentation}
\acro{GENI}{Global Environment for Network Innovations}
\acro{GEYSERS}{Generalized Architecture for Dynamic Infrastructure Services\acroextra{An FP7 EU Project}}
\acro{GLIF}{Global Lambda Integrated Facility}
\acro{GMPLS}{Generalized Multi-Protocol Label Switching}
\acro{IEEE}{Institute of Electrical and Electronics Engineers}
\acro{IETF}{Internet Engineering Task Force}
\acro{IS-IS}{Intermediate System to Intermediate System}
\acro{ITU-T}{Telecommunication Standardization Sector\acroextra{ (coordinates standards on behalf of the \acs{ITU})}}
\acro{ITU}{International Telecommunication Union}
\acro{MIB}{Management Information Base}
\acro{MOMENT}{Monitoring and Measurement Ontology}
\acro{NDL-OWL}{Network Description Language \acs{OWL}}
\acro{NDL}{Network Description Language}
\acro{NM-WG}{Network Measurements Working Group}
\acro{NMC}{Network Measurement and Control WG}
\acro{NML}{Network Markup Language}
\acro{NOVI}{Networking Over Virtualised Infrastructures\acroextra{An FP7 EU Project}}
\acro{OGF}{Open Grid Forum}
\acro{OSPF-TE}{Open Shortest Path First - Traffic Engineering\acroextra{ (An extension of \acs{OSPF})}}
\acro{OSPF}{Open Shortest Path First}
\acro{OWL}{Web Ontology Language}
\acro{PNNI}{Private Network-to-Network Interface}
\acro{RDF}{Resource Description Framework}
\acro{RFC}{Request For Comments\acroextra{ (an \ac{IETF} memorandum on Internet systems and standards)}}
\acro{RIP}{Routing Information Protoocol}
\acro{RSpec}{Resource Specification}
\acro{SFA}{Slice-based Federated Architecture}
\acro{SNMP}{Simple Network Management Protocol}
\acro{VxDL}{Virtual private Execution infrastructure Description Language}
\acro{INDL}{Infrastructure and Network Description Language}
\end{acronym}

\bibliographystyle{plain}
\bibliography{citations,rfc}

\begin{thebibliography}{10}

\bibitem{dmtf}
{Distributed Management Task Force (DMTF)}, 2006.

\bibitem{url:nml}
The network markup language, 2007.

\bibitem{nmwg-schema}
An extensible schema for network measurement and performance data.
\newblock Technical report, Open Grid Forum, 2009.

\bibitem{url:foaf}
Friend of a friend ({FOAF}) project, 2010.

\bibitem{url:fire}
Future internet research and experimentation, 2010.

\bibitem{nmwg}
{OGF Network Measurement Working Group (NMWG)}, 2010.

\bibitem{ogf}
Open grid forum (ogf), 2010.

\bibitem{url:renci}
{RENCI} (renaissance computing institute), 2010.

\bibitem{url:rspec}
Rspec, 2011.

\bibitem{url:SFAv2}
Slice federation architecture, 2011.

\bibitem{cnis-db}
cnis database documentation, 2012.

\bibitem{url:geni}
{GENI} project, 2012.

\bibitem{grid5000}
Grid5000, 2012.

\bibitem{i2-ion}
Internet2 interoperable on-demand network ({ION}), 2012.

\bibitem{lyatiss}
Lyatiss resources, 2012.

\bibitem{url:planetlab}
Planetlab, 2012.

\bibitem{protogeni}
Protogeni, 2012.

\bibitem{esnet}
Energy science network ({ESnet}), 2013.

\bibitem{url:geant}
{G\'EANT} project website, 2013.

\bibitem{autobahn1}
G.~Alyfantis, M.~Balcerkiewicz, J.~Łukasik, R.Krzywania, and G.~Priggouris.
\newblock Technical documentation for the inter- domain bod service manager
  (idm).
\newblock Technical Report DJ3.4.1,2, G\'EANT2, 2007.

\bibitem{orca1}
I.~Baldine, Y.~Xin, D.~Evans, C.~Heerman, J.~Chase, V.~Marupadi, and
  A.~Yumerefendi.
\newblock The missing link: Putting the network in networked cloud computing.
\newblock In {\em International Conference on the Virtual Computing Initiative
  ({ICVCI} 2009)}, October 2009.

\bibitem{perfsonar}
J.W. Boote, E.L. Boyd, J.~Durand, A.~Hanemann, L.~Kudarimoti, R.~{\L}apacz,
  N.~Simar, and S.~Trocha.
\newblock {Towards multi-domain monitoring for the European research networks}.
\newblock {\em Computational Methods in Science and Technology}, 11(2):91--100,
  2005.

\bibitem{rfc5520}
R.~Bradford, JP. Vasseur, and A.~Farrel.
\newblock {Preserving Topology Confidentiality in Inter-Domain Path Computation
  Using a Path-Key-Based Mechanism}.
\newblock RFC 5520 (Proposed Standard), April 2009.

\bibitem{autobahn2}
Mauro Campanella, Radek Krzywania, Afrodite Sevasti, and Stella-Maria Thomas.
\newblock Functional specification and design of a generic domain-centric
  bandwidth on demand service manager.
\newblock Technical Report DJ3.3.4, G\'EANT2, 2008.

\bibitem{rfc3410}
J.~Case, R.~Mundy, D.~Partain, and B.~Stewart.
\newblock {Introduction and Applicability Statements for Internet-Standard
  Management Framework}.
\newblock RFC 3410 (Informational), December 2002.

\bibitem{Cerf1974}
V.~Cerf and R.~Kahn.
\newblock {A Protocol for Packet Network Intercommunication}.
\newblock {\em IEEE Transactions on Communications}, 22(5):637--648, May 1974.

\bibitem{planetlab}
Brent Chun, David Culler, Timothy Roscoe, Andy Bavier, Larry Peterson, Mike
  Wawrzoniak, and Mic Bowman.
\newblock Planetlab: an overlay testbed for broad-coverage services.
\newblock {\em SIGCOMM Comput. Commun. Rev.}, 33(3):3--12, July 2003.

\bibitem{dasgupta07pce}
S.~Dasgupta, J.~C. de~Oliveira, and J.~P. Vasseur.
\newblock {Path-Computation-Element-Based Architecture for Interdomain
  MPLS/GMPLS Traffic Engineering: Overview and Performance}.
\newblock {\em Network, IEEE}, 21(4):38--45, July 2007.

\bibitem{rationale-optical}
Cees de~Laat, Erik Radius, and Steven Wallace.
\newblock The rationale of the current optical networking initiatives.
\newblock {\em Future Generation Computer Systems}, 19(6):999--1008, August
  2003.

\bibitem{idcp}
{DICE}.
\newblock Interdomain controller protocol, 2010.

\bibitem{g805-intro}
Freek Dijkstra, Bert Andree, Karst Koymans, and Jeroen van~der Ham.
\newblock Introduction to {ITU-T} recommendation {G.805}.
\newblock Technical Report UVA-SNE-2007-01, Unversiteit van Amsterdam, December
  2007.

\bibitem{g805-network-model}
Freek Dijkstra, Bert Andree, Karst Koymans, Jeroen van~der Ham, and Cees
  de~Laat.
\newblock A multi-layer network model based on itu-t g.805.
\newblock {\em Computer Networks}, June 2007.

\bibitem{cim}
Distributed Management Task~Force {DMTF}.
\newblock {Common Information Model (CIM)}.

\bibitem{rfc4741}
R.~Enns.
\newblock {NETCONF Configuration Protocol}.
\newblock RFC 4741 (Proposed Standard), December 2006.
\newblock Obsoleted by RFC 6241.

\bibitem{rfc6241}
R.~Enns, M.~Bjorklund, J.~Schoenwaelder, and A.~Bierman.
\newblock {Network Configuration Protocol (NETCONF)}.
\newblock RFC 6241 (Proposed Standard), June 2011.

\bibitem{sfw}
Alberto Escolano, Andrew Mackarel, Damir Regvart, Victor Reijs, Guy Roberts,
  and Hrvoje Popovski.
\newblock Report on testing of technology stitching.
\newblock Technical Report DJ3.5.3, G\'EANT2, 2007.

\bibitem{rfc4655}
A.~Farrel, J.-P. Vasseur, and J.~Ash.
\newblock {A Path Computation Element (PCE)-Based Architecture}.
\newblock RFC 4655 (Informational), August 2006.

\bibitem{farrel-gmpls}
Adrian Farrel and Igor Bryskin.
\newblock {\em {GMPLS}: Architecture and Applications}.
\newblock Morgan Kaufmann, first edition, 2006.

\bibitem{Garcia-Espin2012}
Joan~A. Garcia-Espin, Jordi {Ferrer Riera}, Sergi Figuerola, Mattijs Ghijsen,
  Yuri Demchenko, Jens Buysse, Marc de~Leenheer, Chris Develder, Fabienne
  Anhalt, and Sebastien Soudan.
\newblock {Logical Infrastructure Composition Layer, the GEYSERS Holistic
  Approach for Infrastructure Virtualisation}.
\newblock In {\em Terena Networking Conference (TNC2012)}, 2012.

\bibitem{indl}
M.~Ghijsen, J.~van~der Ham, P.~Grosso, and C.~de~Laat.
\newblock {Towards an Infrastructure Description Language for Modeling
  Computing Infrastructures}.
\newblock In {\em Parallel and Distributed Processing with Applications (ISPA),
  2012 IEEE 10th International Symposium on}, pages 207--214. IEEE, July 2012.

\bibitem{indl2}
M.~Ghijsen, J.~van~der Ham, P.~Grosso, Cosmin Dumitru, Hao Zhu, Zhiming Zhao,
  and C.~de~Laat.
\newblock A semantic-web approach for modeling computing infrastructures.
\newblock Technical Report UVA-SNE-2013-01, University of Amsterdam, SNE group,
  March 2013.

\bibitem{guok06oscars}
Chin Guok, D.~Robertson, M.~Thompson, J.~Lee, B.~Tierney, and W.~Johnston.
\newblock {Intra and Interdomain Circuit Provisioning Using the OSCARS
  Reservation System}.
\newblock In {\em Broadband Communications, Networks and Systems, 2006.
  BROADNETS 2006. 3rd International Conference on}, pages 1--8. IEEE, October
  2006.

\bibitem{perfsonar2}
Andreas Hanemann, Jeff Boote, Eric Boyd, J\'er\^ome Durand, Loukik Kudarimoti,
  Roman \L{}apacz, Martin Swany, Szymon Trocha, and Jason Zurawski.
\newblock Perfsonar: A service oriented architecture for multi-domain network
  monitoring.
\newblock In Boualem Benatallah, Fabio Casati, and Paolo Traverso, editors,
  {\em Service-Oriented Computing - ICSOC 2005}, volume 3826 of {\em Lecture
  Notes in Computer Science}, pages 241--254. Springer Berlin / Heidelberg,
  2005.
\newblock 10.1007/11596141\_19.

\bibitem{Heart1970}
F.~E. Heart, R.~E. Kahn, S.~M. Ornstein, W.~R. Crowther, and D.~C. Walden.
\newblock {The interface message processor for the ARPA computer network}.
\newblock In {\em Proceedings of the May 5-7, 1970, spring joint computer
  conference on - AFIPS '70 (Spring)}, page 551, New York, New York, USA, May
  1970. ACM Press.

\bibitem{rfc1058}
C.L. Hedrick.
\newblock {Routing Information Protocol}.
\newblock RFC 1058 (Historic), June 1988.
\newblock Updated by RFCs 1388, 1723.

\bibitem{ss7-routing}
International Telecommunication~Union ({ITU}).
\newblock Signalling network functions and messages.
\newblock Recommendation {ITU-T} {Q.704}, International Telecommunication Union
  ({ITU}), July 1996.

\bibitem{g805}
International Telecommunication~Union ({ITU}).
\newblock Generic functional architecture of transport networks.
\newblock Recommendation {ITU-T} {G.805}, International Telecommunication Union
  ({ITU}), March 2000.

\bibitem{g809}
International Telecommunication~Union ({ITU}).
\newblock Functional architecture of connectionless layer networks.
\newblock Recommendation {ITU-T} {G.809}, International Telecommunication Union
  ({ITU}), 2003.

\bibitem{g800}
International Telecommunication~Union ({ITU}).
\newblock Unified functional architecture of transport networks.
\newblock Recommendation {ITU-T} {G.800}, International Telecommunication Union
  ({ITU}), 2007.

\bibitem{vxdl}
Guilherme~Piegas Koslovski, Pascale Vicat-Blanc Primet, and Andrea~Schwertner
  Charão.
\newblock Vxdl: Virtual resources and interconnection networks description
  language.
\newblock In {\em Networks for Grid Applications}, volume~2 of {\em Lecture
  Notes of the Institute for Computer Sciences, Social Informatics and
  Telecommunications Engineering}, pages 138--154. Springer Berlin Heidelberg,
  2009.

\bibitem{Kuipers200978}
Fernando Kuipers and Freek Dijkstra.
\newblock Path selection in multi-layer networks.
\newblock {\em Computer Communications}, 32(1):78 -- 85, 2009.

\bibitem{lehman11multilayer}
T.~Lehman, Xi~Yang, N.~Ghani, Feng Gu, Chin Guok, I.~Monga, and B.~Tierney.
\newblock Multilayer networks: an architecture framework.
\newblock {\em Communications Magazine, IEEE}, 49(5):122 --130, may 2011.

\bibitem{rfc1105}
K.~Lougheed and Y.~Rekhter.
\newblock {Border Gateway Protocol (BGP)}.
\newblock RFC 1105 (Experimental), June 1989.
\newblock Obsoleted by RFC 1163.

\bibitem{rfc3945}
E.~Mannie.
\newblock {Generalized Multi-Protocol Label Switching (GMPLS) Architecture}.
\newblock RFC 3945 (Proposed Standard), October 2004.
\newblock Updated by RFC 6002.

\bibitem{moment}
Peter M\'atray, I.~Csabai, and P.~H\'aga.
\newblock A semantic extension of the network measurement virtual observatory.
\newblock 2009.

\bibitem{Matray07etomic}
Peter Matray, Istvan Csabai, Peter Haga, Jozsef Steger, Laszlo Dobos, and Gabor
  Vattay.
\newblock Building a prototype for network measurement virtual observatory.
\newblock In {\em Proceedings of the 3rd annual ACM workshop on Mining network
  data}, MineNet '07, pages 23--28, New York, NY, USA, 2007. ACM.

\bibitem{McQuillan1979}
John~M. McQuillan, Ira Richer, and Eric~C. Rosen.
\newblock {An overview of the new routing algorithm for the ARPANET}.
\newblock In {\em Proceedings of the sixth symposium on Data communications -
  SIGCOMM '79}, pages 63--68, New York, New York, USA, November 1979. ACM
  Press.

\bibitem{rfc4274}
D.~Meyer and K.~Patel.
\newblock {BGP-4 Protocol Analysis}.
\newblock RFC 4274 (Informational), January 2006.

\bibitem{Perlman_1991}
Radia Perlman.
\newblock A comparison between two routing protocols: Ospf and is-is.
\newblock {\em Ieee Network}, 5(5):18--24, 1991.

\bibitem{pnnispec}
Private network-network interface specification.
\newblock Technical report, {ATM Forum}, 1996.

\bibitem{nsiv1}
Guy Roberts, Tomohiro Kudoh, Inder Monga, Jerry Sobieski, and John Vollbrecht.
\newblock {Network Services Framework v1.0 }, March 2010.

\bibitem{arpanet}
Peter~H. Salus.
\newblock {\em Casting the Net: From ARPANET to Internet and Beyond...}
\newblock Addison-Wesley Longman Publishing Co., Inc., Boston, MA, USA, 1995.

\bibitem{den-ng}
J.~Strassner.
\newblock {DEN}-ng: achieving business-driven network management.
\newblock In {\em NOMS 2002. IEEE/IFIP Network Operations and Management
  Symposium. ' Management Solutions for the New Communications World'(Cat.
  No.02CH37327)}, pages 753--766. IEEE, August 2002.

\bibitem{NOVID22}
Jeroen van~der Ham, Mauro Campanella, Alejandro Chuang, Fabio Farina, Paola
  Grosso, Yiannos Kryftis, P\'eter M\'atray, Alvaro Monje, Chrysa Papagianni,
  Chariklis Pittaras, Celia~Velayos J\'ozsef~St\'eger, Adianto Wibisono, and
  Klaas Wierenga.
\newblock D2.2: First information and data models.
\newblock Technical report, {NOVI} Consortium, October 2011.

\bibitem{vanderHam200885}
Jeroen van~der Ham, Freek Dijkstra, Paola Grosso, Ronald van~der Pol, Andree
  Toonk, and Cees de~Laat.
\newblock A distributed topology information system for optical networks based
  on the semantic web.
\newblock {\em Optical Switching and Networking}, 5(2–3):85 -- 93, 2008.
\newblock <ce:title>Advances in IP-Optical Networking for IP Quad-play Traffic
  and Services</ce:title>.

\bibitem{vdham06ndl}
Jeroen van~der Ham, Freek Dijkstra, Franco Travostino, Hubertus Andree, and
  Cees de~Laat.
\newblock Using rdf to describe networks.
\newblock {\em Future Generation Computer Systems, Feature topic iGrid 2005},
  October 2006.

\bibitem{nml-schema}
Jeroen van~der Ham, Freek Dijkstra, Roman Łapacz, and Jason Zurawski.
\newblock {Network Markup Language Base Schema version 1}, June 2013.

\bibitem{VanderHam2011}
Jeroen van~der Ham, Chrysa Papagianni, Jozsef Steger, Peter Matray, Yiannos
  Kryftis, Paola Grosso, and Leonidas Lymberopoulos.
\newblock {Challenges of an information model for federating virtualized
  infrastructures}.
\newblock In {\em 5th International DMTF Academic Alliance Workshop on Systems
  and Virtualization Management: Standards and the Cloud}, 2011.

\bibitem{orca2}
Yufeng Xin, I.~Baldine, J.~Chase, T.~Beyene, B.~Parkhurst, and A.~Chakrabortty.
\newblock Virtual smart grid architecture and control framework.
\newblock In {\em Smart Grid Communications (SmartGridComm), 2011 IEEE
  International Conference on}, pages 1 --6, oct. 2011.

\bibitem{perfsonar-inter-domain}
M.~Yampolskiy and M.K. Hamm.
\newblock Management of multidomain end-to-end links - a federated approach for
  the pan-european research network geant 2.
\newblock In {\em Integrated Network Management, 2007. IM '07. 10th IFIP/IEEE
  International Symposium on}, pages 189 --198, may 2007.

\bibitem{snmp-book}
Dave Zeltserman.
\newblock {\em Practical Guide to {SNMPv3} and Network Management}.
\newblock Prentice Hall PTR, 1999.

\end{thebibliography}
\end{document}